\documentstyle[twocolumn,graphicx]{jpsj}
\newcommand{\eps}{\varepsilon}

\newcommand{\vct}[1]{\mbox{\boldmath #1}}
\def\runtitle{Spin and Orbital States 
and Their Phase Transitions of the Perovskite-Type Ti Oxides:
Weak coupling Approach}
\def\runauthor{Masahito {\sc Mochizuki}}
\title{Spin and Orbital States and Their Phase Transitions 
of the Perovskite-Type Ti Oxides: Weak Coupling Approach}
\author
{Masahito {\sc Mochizuki}}

\inst
{Institute for Solid State Physics, 
University of Tokyo,\\ 5-1-5 Kashiwa-no-ha, Kashiwa, Chiba 277-8581}

\recdate{\today}

\abst{The magnetic phase diagram of the 
perovskite-type Ti oxides as a function of the
${\rm GdFeO}_3$-type distortion is examined by using the
Hartree-Fock analysis of a  
multiband $d$-$p$ Hamiltonian from a viewpoint
of competitions of the spin-orbit interaction, 
the Jahn-Teller (JT) level-splitting and spin-orbital 
superexchange interactions.
Near the antiferromagnetic (AFM)-to-ferromagnetic (FM) phase boundary, 
A-type AFM [AFM(A)] and FM states accompanied by a certain
type of orbital ordering are lowered in energy
at large JT distortion, which is in agreement with 
the previous strong coupling study.
With increasing the ${\rm GdFeO}_3$-type distortion,
their phase transition occurs.
Through this magnetic phase transition, 
the orbital state hardly changes, which induces nearly continuous 
change in the spin coupling along the $c$-axis
from negative to positive.
The resultant strong two-dimensionality in the spin coupling
near the phase boundary
is attributed to the strong suppression of $T_{\rm N}$ and $T_{\rm C}$,
which is experimentally observed.
On the other hand, at small ${\rm GdFeO}_3$-type without
JT distortions, which correspond to
${\rm LaTiO}_3$, the most stable solution is not G-type 
AFM [AFM(G)] but FM. 
Although the spin-orbit interaction has been considered to be 
relevant at the small or no JT distortion of ${\rm LaTiO}_3$ 
in the literature, our analysis
indicates that the spin-orbit interaction
is irrelevant to the AFM(G) state in ${\rm LaTiO}_3$
and superexchange-type interaction dominates.
On the basis of further investigations on the nature of this
FM state and other solutions, this discrepancy is discussed
in detail.}

\kword
{perovskite-type Ti oxides, ${\rm GdFeO}_3$-type distortion, 
$d$-level degeneracy, $d$-type Jahn-Teller distortion, 
spin-orbit interaction, multiband $d$-$p$ model}
\begin{document}
\sloppy
\maketitle
\section{Introduction}

In transition-metal oxides, strong electron correlations often 
localize the $3d$ electrons and the system 
becomes an insulator (a Mott insulator).~\cite{Imada98}
These compounds have recently attracted considerable
interest since they show rich magnetic and orbital phases.
In particular, perovskite-type oxides $RM{\rm O}_3$,
where $R$ denotes a trivalent rare-earth ion
(i.e., La, Pr, Nd, ..., Y) and $M$ is a transition-metal 
ion (i.e., Ti, V, ..., Ni, Cu) exhibit a variety
of magnetic and electronic properties caused by an interplay 
of charge, spin and orbital degrees of freedom. 

The perovskite-type Ti oxide $R{\rm TiO}_3$ is a prototypical example.
In these compounds, ${\rm Ti}^{3+}$ has a $t_{2g}^1$ 
configuration, and one of the threefold 
$t_{2g}$-orbitals is occupied at each transition-metal site. 
They have attracted much interest since these systems show
various magnetic and orbital orderings owing to the threefold 
degeneracy of the $t_{2g}$ orbitals. 
It requires to take both spin and orbital 
fluctuations into consideration to explain competitions 
of such rich phases.
Moreover, the spin-orbit interaction would make the magnetic 
and orbital structures more complicated
since the $t_{2g}$ orbitals are strongly affected by the interaction.

\begin{figure}[tdp]
\hfil
\includegraphics[scale=0.45]{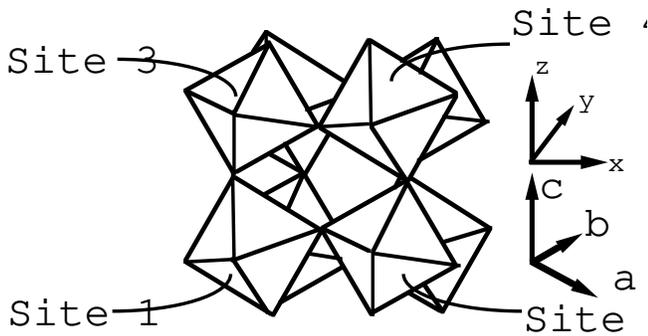}
\hfil
\caption{${\rm GdFeO}_3$-type distortion.}
\label{gdfo3}
\end{figure}

The crystal structure of $R{\rm TiO}_6$ is an orthorhombically
distorted cubic-perovskite (${\rm GdFeO}_3$-type distortion)
in which the ${\rm TiO}_6$ octahedra forming the perovskite 
lattice tilt alternatingly as shown in Fig.~\ref{gdfo3}. 
The magnitude of the distortion depends on the ionic radii
of the $R$ ions. With a small ionic radius of the $R$ ion, 
the lattice structure is more distorted and the bond angle
is more significantly decreased from $180^{\circ}$.
In ${\rm LaTiO}_3$, the bond angle is 
$157^{\circ}$ ($ab$-plane)
and $156^{\circ}$ ($c$-axis), but $144^{\circ}$ ($ab$-plane) 
and $140^{\circ}$ ($c$-axis) in ${\rm YTiO}_3$~\cite{MacLean79}.
The distortion can be controlled by the use of the solid-solution systems  
${\rm La}_{1-y}{\rm Y}_{y}{\rm TiO}_3$
or in $R{\rm TiO}_3$, by varying the $R$ ions.
In particular, by varying the Y concentration in 
${\rm La}_{1-y}{\rm Y}_{y}{\rm TiO}_3$, we can control the 
bond angle almost continuously from $157^{\circ}$ ($y=0$) 
to $140^{\circ}$ ($y=1$). 

In ${\rm YTiO}_3$, a $d$-type JT distortion has been observed 
in which the longer and shorter Ti-O bond lengths are 
$\sim$2.08 $\AA$ and $\sim$2.02 $\AA$, respectively.~\cite{Akimitsu98}
In the $d$-type JT distortion, the $xy$ and $yz$ orbitals are stabilized
at sites 1 and 3, and the $xy$ and $zx$ orbitals are stabilized
at sites 2 and 4. 
On the other hand, ${\rm LaTiO}_3$ exhibits no detectable JT distortion.

Recently, the magnetic phase diagrams have been studied
as functions of the magnitude of ${\rm GdFeO}_3$-type
distortion.~\cite{Goral82,Greedan85,Okimoto95,Katsufuji97}
In La-rich ($y<0.6$) systems or in the compounds with large
$R$ ions, in which the ${\rm GdFeO}_3$-type distortion 
is relatively small, an AFM ground state is realized.
In particular, ${\rm LaTiO}_3$ exhibits a AFM(G) ground state
with magnetic moment of 0.45 $\mu_{\rm B}$, which 
is strongly reduced from spin-only moment, and the
N${\rm {\grave{e}}el}$ temperature ($T_{\rm N}$) is about 130 K.
With increasing the Y concentration or varying the $R$ site with 
smaller-sized ions (an increase of the ${\rm GdFeO}_3$-type distortion),
$T_{\rm N}$ decreases rapidly and is suppressed to almost zero,
subsequently a FM ordering appears.
In Y-rich systems and in ${\rm YTiO}_3$ in which the 
${\rm GdFeO}_3$-type distortion is relatively large, the system shows
a FM ground state.

In order to elucidate these phase diagrams, model Hartree-Fock studies 
have been done previously.~\cite{Mizokawa95,Mizokawa96a}
In these weak coupling studies, it is claimed that in 
${\rm LaTiO}_3$ with small ${\rm GdFeO}_3$-type distortion, 
a AFM(G) state with the spin-orbit ground state is realized
for the small or no JT distortion, and resultant unquenched
orbital moment is considered to be consistent with the strong
reduction of the moment.
On the other hand, a FM state accompanied by an orbital 
ordering is realized in ${\rm YTiO}_3$ with large JT distortion.
However, in these studies, the nature of the phase
diagrams has not been elucidated sufficiently in the following sense.
At first sight, we can expect the first-order transition 
between completely different symmetry breaking in which
$T_{\rm N}$ and $T_{\rm C}$ remain nonzero at the AFM-FM phase boundary.
However, in the magnetic phase diagrams, 
$T_{\rm N}$ and $T_{\rm C}$ are strongly suppressed around the phase boundary.
This strong suppression implies a continuous-type transition 
at $T=0$ and contradicts our naive expectation. 
This second-order like phase transition is not explained 
in these studies, and has been an issue of interest.

Recently, in order to clarify this problem, 
effective Hamiltonian in the insulating limit has been applied
to this system.~\cite{Mochizuki00,Mochizuki01a}
According to these strong coupling studies, in the AFM phase near 
the AFM-FM phase boundary, an AFM(A) ground state is realized.
This AFM(A) phase has not been studied in the previous 
weak coupling approach.  
Moreover, since the orbital state is strongly stabilized and 
changes only little through the transition, strong two-dimensionality
in spin-coupling is predicted near the phase boundary and the
strong suppressions of $T_{\rm N}$ and $T_{\rm C}$
are naturally understood. In these studies, 
a large JT distortion is assumed in order to focus on the situation
near the AFM-FM phase boundary, and the spin-orbit interaction is neglected
on the basis of the large energy-splitting due to the JT distortion.
In addition, the AFM(G) state in ${\rm LaTiO}_3$ with small or no 
JT distortion has not been reproduced.

However, the spin-orbit interaction may become relevant
in the systems with small or no JT distortion such as ${\rm LaTiO}_3$.
While the spin-orbit interaction can be neglected if the JT 
distortion is large, we can expect a strong competition 
between the spin-orbit interaction and JT level-splitting with 
decreasing the JT distortion.
With sufficiently small JT distortion, 
the system may well be described by the spin-orbit ground state. 
Besides, successive spin-orbital superexchange interactions
may dominate over the spin-orbit interaction even without JT distortion.
At this stage, it is an issue of importance to examine the phase
diagrams from a viewpoint of their competitions.
In the weak coupling approach
in which transfers of electrons and spin-orbit
interaction are treated in a non-perturbative manner,
we can consider both effects on an equal footing.
This approach is appropriate for a systematic study on the interplay of them. 
In these senses, the weak-coupling and the strong-coupling studies 
are complementary to each other, and analysis from the weak 
coupling approach is important.

In this paper, we investigate the magnetic phase diagrams
by using the Hartree-Fock analysis of the multiband $d$-$p$ 
Hamiltonian.
We study the magnetic and orbital states as functions of the 
${\rm GdFeO}_3$-type and $d$-type JT distortions. 
Since effects of both electron transfers and spin-orbit interaction 
are taken into account on an equal footing, this model
is appropriate for a study on the competitions of  
spin-orbit interaction, JT level-splitting and 
superexchange interactions.
The weak coupling treatment does not properly reproduce the energy 
scale of the superexchange interaction $J$ defined in the strong 
coupling region, where $J$ is proportional to $t^2/U$ with 
$t$ and $U$ being typical transfer and on-site Coulomb repulsion.
However, the physics contained in the reproduction of the
superexchange interaction with AFM and/or antiferro-orbital 
(AF-orbital) is expected to be adiabatically 
connected with the SDW type symmetry breaking in the
weak-coupling Hartree-Fock solution.
Therefore, we will refer the stabilization of the SDW
(or orbital density wave) type  
solution with AFM (or AF-orbital) symmetry breaking to the 
superexchange mechanism.

Pioneering works by using this method have already 
been done by Mizokawa and Fujimori.~\cite{Mizokawa95,Mizokawa96a}
However, concerning the region of small ${\rm GdFeO}_3$-type distortion,
we have come to a different conclusion
by studying orbital-spin states which they have overlooked.
We show that in the small ${\rm GdFeO}_3$-type distortion
without JT level-splitting, a FM spin state accompanied by an 
AF-orbital ordering is stabilized by the energy gains of both 
spin-orbit and superexchange interactions.
In this FM solution, the spin-orbit ground state is not realized
at certain sites, which suggests that 
the spin-orbital superexchange interactions
due to the electron transfers dominate over  
the spin-orbit interaction even without JT distortion.
In the previous studies, AFM(G) state with spin-orbit ground state 
has been claimed to be stabilized without JT distortion.
However, in these studies, the stabilization of this AFM(G) state
is concluded only from comparison of the energies between
this AFM(G) solution and a FM solution with higher energy,
and our FM solution is ignored.
Our FM state has not been reproduced so far, and is studied 
for the first time by our weak coupling approach.
We conclude that the AFM(G) state 
in ${\rm LaTiO}_3$ does not accompany with spin-orbit ground state,
and there exists another origin for its emergence. 
Recent neutron-scattering experiment
shows the spin-wave spectrum of ${\rm LaTiO}_3$ 
well described by a spin-1/2 isotropic Heisenberg model
on the cubic lattice, and absence of unquenched 
orbital momentum.~\cite{Keimer00}
This result also seems to contradict the naive prediction of spin-orbit 
ground state with no JT distortion. 
By studying a model including the spin-orbit interaction,
we propose some statements on this experimental result.

Moreover, we apply this method to the systems near the AFM-FM phase  
boundary for the first time.
We show that the strange behavior of the magnetic phase transition 
is well described on the basis of JT ground state when we consider 
the experimentally observed large JT distortion.
The results on the properties and nature of the phase 
transition are in agreement with those obtained by the previous  
strong coupling approaches, which indicates its validity irrespective 
of the coupling strength.~\cite{Mochizuki00,Mochizuki01a}
In addition, we study a magnetic and orbital phase diagram
in the plane of the ${\rm GdFeO}_3$-type and $d$-type JT distortions
in order to examine how extent the physics of AFM-FM phase transition
in strong coupling limit survives when the JT level-splitting
competes with the spin-orbit interaction.

The organization of this paper is as follows.
In $\S$ 2, we introduce the multiband $d$-$p$ Hamiltonian
to describe the realistic systems of the perovskite-type Ti oxides.
In $\S$ 3, numerical results calculated by applying 
the unrestricted Hartree-Fock approximation are presented.
Section. 4 is devoted to the summary and conclusions.

\section{Multiband $d$-$p$ model}

 We employ the following Hamiltonian:
\begin{equation}
        H^{dp} = H_{d0} + H_{p} + H_{tdp} + H_{tpp}
               + H_{h} + H_{\rm on-site}, \\
\label{dph}
\end{equation}
with
\begin{eqnarray}
     & &H_{d0} = \sum_{\alpha,i,\gamma,\sigma} \eps_{d}^0
    d_{\alpha,i \gamma \sigma}^{\dagger} d_{\alpha,i \gamma \sigma}, \\
     & &H_{p} = \sum_{\alpha,j,l,\sigma} \eps_{p}
    p_{\alpha,j l \sigma}^{\dagger} p_{\alpha,j l \sigma}, \\
     & &H_{tdp} = \sum_{\alpha,i,\gamma,\alpha',j,l,\sigma} 
        t_{\alpha i\gamma,\alpha'jl}^{dp}
                      d_{\alpha,i \gamma \sigma}^{\dagger} 
                      p_{\alpha',j l \sigma}  + \vct{h.c.}, \\
     & &H_{tpp} = \sum_{\alpha,j,l,\alpha',j',l',\sigma} 
        t_{\alpha jl,\alpha'j'l'}^{pp}
                      p_{\alpha,j l \sigma}^{\dagger} 
                      p_{\alpha',j' l' \sigma}  + \vct{h.c.}, \\
     & &H_{h} = \sum_{\alpha,i,\gamma,\gamma',\sigma,\sigma'}
       h_{\gamma\sigma,\gamma'\sigma'}
        d_{\alpha,i \gamma \sigma}^{\dagger} d_{\alpha,i \gamma' \sigma'}, \\
     & &H_{\rm on-site} = H_{u} + H_{u'} + 
                            H_j + H_{j'}, 
\label{dph2}
\end{eqnarray}  
where ${d_{\alpha,i \gamma \sigma}^{\dagger}}$ 
is a creation operator of an electron 
with spin $\sigma(={\uparrow},{\downarrow})$ in the
$3d$ orbital $\gamma$ at Ti site $i$ 
in the $\alpha$-th unit cell, 
and ${p_{\alpha,j l \sigma}^{\dagger}}$ 
is a creation operator of an electron 
with spin $\sigma(={\uparrow},{\downarrow})$ in the
$2p$ orbital $l$ at oxygen site $j$ in the $\alpha$-th unit cell.
Here, $H_{d0}$ and $H_p$ stand for the bare level energies of Ti $3d$ and  
O $2p$ orbitals, respectively.
$H_{tdp}$ and $H_{tpp}$ are $d$-$p$ and $p$-$p$ hybridization
terms, respectively.
$H_{h}$ denotes the crystal field and spin-orbit interaction represented
by the parameter $\xi=0.018$ eV.~\cite{Sugano70} 
The term $H_{\rm on-site}$ represents on-site $d$-$d$ Coulomb interactions.
${t^{dp}_{\alpha i\gamma,\alpha'jl}}$ and
${t^{pp}_{\alpha jl,\alpha'{j^{\prime}}{l^{\prime}}}}$ 
are nearest-neighbor $d$-$p$ and $p$-$p$ transfers, respectively,
which are given in terms of 
Slater-Koster parameters $V_{pd{\pi}}$, $V_{pd{\sigma}}$, $V_{pp{\pi}}$ 
and $V_{pp{\sigma}}$.~\cite{Slater54}
$H_{\rm on-site}$ term consists of the following 
four contributions:
\begin{eqnarray}
     & &   H_{u} = \sum_{\alpha,i,\gamma} u
        d_{\alpha,i \gamma \uparrow}^{\dagger} 
        d_{\alpha,i \gamma \uparrow}
        d_{\alpha,i \gamma \downarrow}^{\dagger} 
        d_{\alpha,i \gamma \downarrow}, \\
     & &   H_{u'} = \!\! 
        \sum_{\alpha,i,\gamma>\gamma',{\sigma},{\sigma}'} u'
        d_{\alpha,i \gamma \sigma}^{\dagger} 
        d_{\alpha,i \gamma \sigma}
        d_{\alpha,i \gamma' {\sigma}'}^{\dagger}
        d_{\alpha,i \gamma' {\sigma}'}, \\
     & &   H_{j} = \!\!   
        \sum_{\alpha,i,\gamma>\gamma'\sigma,{\sigma}'} j
        d_{\alpha,i \gamma \sigma}^{\dagger} 
        d_{\alpha,i \gamma' \sigma}
        d_{\alpha,i \gamma' {\sigma}'}^{\dagger} 
        d_{\alpha,i \gamma {\sigma}'}, \\
     & &   H_{j'} = \sum_{\alpha,i,\gamma \ne \gamma'} j'
        d_{\alpha,i \gamma \uparrow}^{\dagger} 
        d_{\alpha,i \gamma' \uparrow}
        d_{\alpha,i \gamma \downarrow}^{\dagger} 
        d_{\alpha,i \gamma' \downarrow}, 
\end{eqnarray}
where $H_{u}$ and $H_{u'}$ are the intra- and inter-orbital 
Coulomb interactions and $H_{j}$ and $H_{j'}$ denote the
exchange interactions.
The term $H_{j}$ is the origin of the Hund's rule coupling 
which strongly favors the spin alignment in the same
direction on the same atoms.
These interactions are expressed by Kanamori parameters,
$u$, $u^{\prime}$, $j$ and $j^{\prime}$ which 
satisfy the following relations:~\cite{Brandow77,Kanamori63}
\begin{eqnarray}
     u &=& U + \frac{20}{9}j , \\
     u'&=& u -2j , \\
     j &=& j'.
\end{eqnarray}
Here, $U$ gives the magnitude of the multiplet-averaged 
$d$-$d$ Coulomb interaction. 
The charge-transfer energy $\Delta$, which describes the energy
difference between occupied O $2p$ and unoccupied
Ti $3d$ orbitals, is
defined by $U$ and energies of the
bare Ti $3d$ and O $2p$ orbitals $\eps_d^0$ and $\eps_p$
as follows,
\begin{equation}
      \Delta = \eps_{d}^0 + U -\eps_p.
\end{equation}
The values of $\Delta$, $U$ and $V_{pd\sigma}$ are
estimated by the cluster-model analyses of valence-band and
transition-metal $2p$ core-level photoemission 
spectra.~\cite{Saitoh95,Bocquet96}
We take the values of these parameters as
$\Delta = 7.0$ eV, $U = 4.0$ eV, $V_{pd\sigma} = -2.2$ eV
and $j = 0.64$ eV
throughout the present calculation. 
The ratio $V_{pd\sigma}/V_{pd\pi}$ is fixed at 
$-2.18$, and $V_{pp\sigma}$ and $V_{pp\pi}$ at 0.60 eV 
and $-0.15$ eV, respectively~\cite{Harrison89,Mattheiss72a,Mattheiss72b}.
The effects of the ${\rm GdFeO}_3$-type distortion and the $d$-type 
JT distortion are reflected on the 
hopping integrals.
The ${\rm GdFeO}_3$-type structure is orthorhombic with orthogonal
$a$-, $b$- and $c$-axes which can be obtained by rotating the four 
octahedra in the unit cell.
Let us represent the four octahedra in the unit cell 
as site 1, site 2, site 3 and site 4 as shown in Fig.~\ref{gdfo3}.
Here, we simulate the ${\rm GdFeO}_3$-type structure by tilting 
the ${\rm TiO}_6$ octahedra  by $+\theta$ and
$-{\theta}$ about the (1,1,1) and ($-1,-1,1$) axes with respect to the
$x$,$y$ and $z$ axes.
The magnitude of the ${\rm GdFeO}_3$-type distortion is
expressed by the bond angle.
The magnitude of the JT distortion can be denoted
by the ratio $[V_{pd{\sigma}}^s$/$V_{pd{\sigma}}^l]^{1/3}$;
here, $V_{pd{\sigma}}^s$ and $V_{pd{\sigma}}^l$
are the transfer integrals for the shorter and longer
Ti-O bonds, respectively.
The value for ${\rm YTiO}_3$ estimated by using Harrison's
rule takes $\sim$1.040~\cite{Harrison89}.
This large JT distortion is also considered to be realized near the 
AFM-FM phase boundary.

We can rewrite the Hamiltonian in the $\vct{k}$-space form
by using the following Bloch-electron operators,
\begin{eqnarray}
d_{\vct{$k$},i \gamma \sigma}^{\dagger} &=& 
\frac{1}{\sqrt{N}}\sum_{\alpha} e^{i \vct{$k$}\cdot{\vct{$R$}_{\alpha}}}
d_{\alpha,i \gamma \sigma}^{\dagger},          \\
p_{\vct{$k$},j l \sigma}^{\dagger} &=& 
\frac{1}{\sqrt{N}}\sum_{\alpha} e^{i \vct{$k$}\cdot{\vct{$R$}_{\alpha}}}
p_{\alpha,j l \sigma}^{\dagger},               
\end{eqnarray}  
where $\vct{$k$}$ labels the wave vector in the first
Brillouin zone.

\section{Results and Discussions}

In this section, we present the numerical results
calculated by applying the unrestricted Hartree-Fock approximation 
to the multiband $d$-$p$ model introduced in the previous section.
In our calculations, we have concentrated on uniform solutions. 
At this stage, the order parameters can be written as,
\begin{equation}
\langle d_{\alpha,i \gamma \sigma}^{\dagger}
d_{\alpha,i \gamma' \sigma'} \rangle = 
\frac{1}{N} \sum_{\vct{$k$}}
\langle d_{\vct{$k$},i \gamma \sigma}^{\dagger}
d_{\vct{$k$},i \gamma' \sigma'} \rangle
\end{equation}
which are to be determined self-consistently.
We have taken 512$\vct{$k$}$ points in the first Brillouin zone
of the ${\rm GdFeO}_3$-type structure and iterated the self-consistency
cycle until the convergence of all the order parameters
within errors of $1 \times 10^{-4}$.
It should be noted that the basis of the Ti $3d$ orbitals
are defined by using $x$-, $y$-, and $z$-axes attached to each 
${\rm TiO}_6$ octahedron in this paper.
 
First, in order to focus on the situation 
near the AFM-FM phase boundary,
the magnitude of the JT distortion: 
$[V_{pd{\sigma}}^s$/$V_{pd{\sigma}}^l]^{1/3}$
is fixed at 1.040, which is considered to 
be realized around the AFM-FM phase boundary.

\begin{figure}[tdp]
\hfil
\includegraphics[scale=0.5]{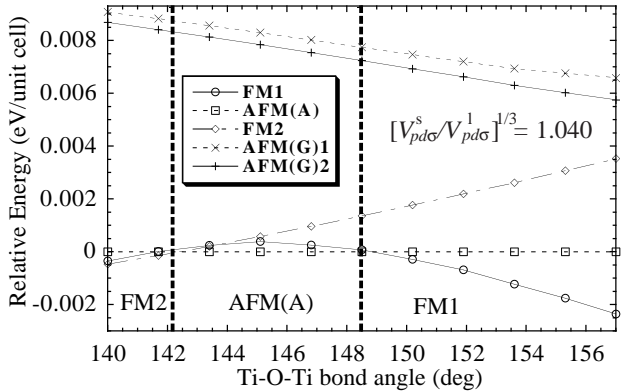}
\hfil
\caption{Energies of various spin and orbital configurations
relative to that of AFM(A) state
as functions of the Ti-O-Ti bond angle in the case 
of $[V_{pd{\sigma}}^s$/$V_{pd{\sigma}}^l]^{1/3}=1.040$.}
\label{re1.040}
\end{figure}
In Fig.~\ref{re1.040}, relative energies of various 
spin and orbital configurations are plotted as functions 
of the Ti-O-Ti bond angle
from $157^{\circ}$ to $140^{\circ}$.
\begin{figure}[tdp]
\hfil
\includegraphics[scale=0.6]{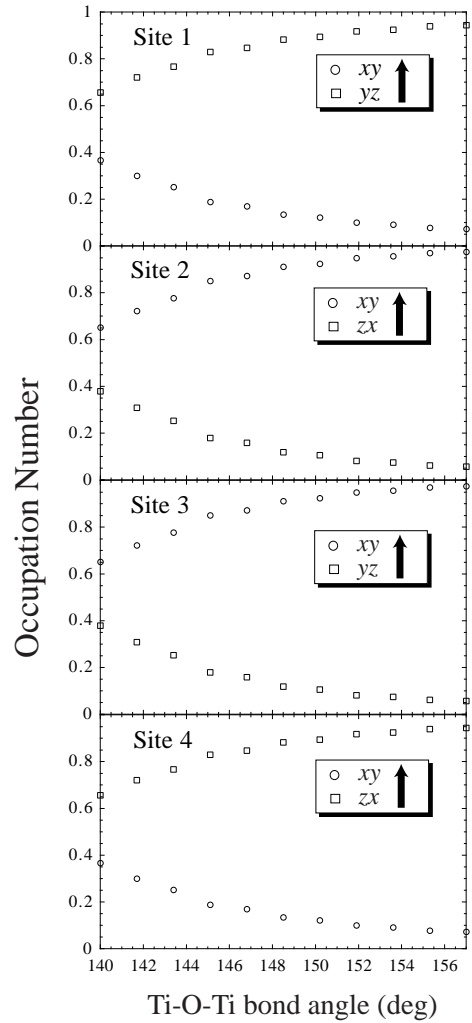}
\hfil
\caption{The orbital occupation in the majority spin states
of FM1 state as a function of the Ti-O-Ti bond angle in the case 
of $[V_{pd{\sigma}}^s$/$V_{pd{\sigma}}^l]^{1/3}=1.040$.}
\label{edns_fm1}
\end{figure}
In the small ${\rm GdFeO}_3$-type 
distortion, a FM solution with ($yz,xy,xy,zx$)-type orbital
ordering in which site 1, 2, 3 and 4 are dominantly occupied by 
$yz$, $xy$, $xy$ and $zx$, respectively 
(FM1 solution) is stabilized (see Fig.~\ref{edns_fm1})
since the FM state with the orbital configuration
in which the neighboring occupied-orbitals are approximately
orthogonal (AF-orbital ordering) is favored both by transfers 
and by the exchange interaction $j$.  
A AFM(G) solution with ($yz,xy,xy,zx$)-type orbital ordering [AFM(G)1]
has much higher energy.
However, it should be noted that
the present calculations are carried out in the case of 
large JT distortion so that the obtained FM1 solution with the small
${\rm GdFeO}_3$-type distortion does not necessarily contradict the 
emergence of AFM(G)-ground state in ${\rm LaTiO}_3$
with no JT distortion. 

\begin{figure}[tdp]
\hfil
\includegraphics[scale=0.5]{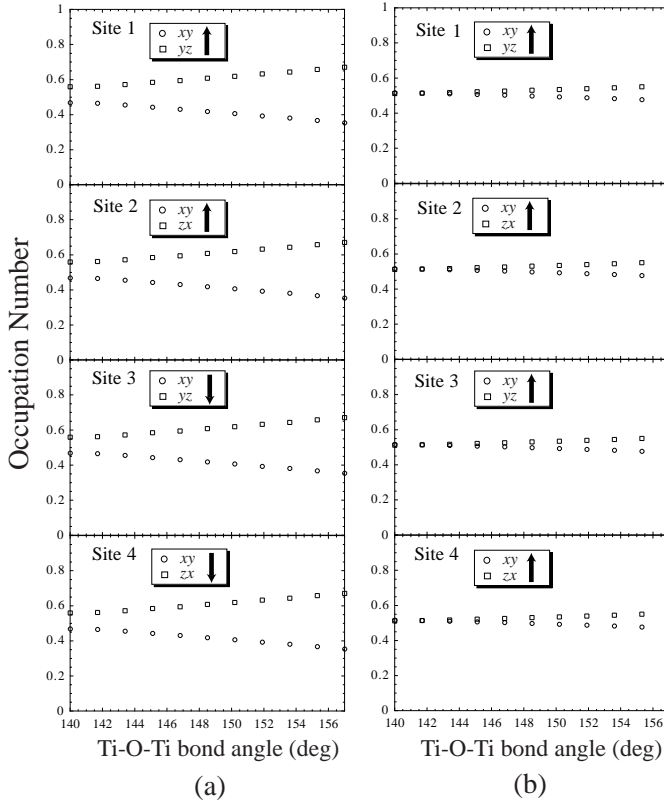}
\hfil
\caption{The orbital occupation in the majority spin states
of (a) AFM(A) and (b) FM2 states
as functions of the Ti-O-Ti bond angle in the case 
of $[V_{pd{\sigma}}^s$/$V_{pd{\sigma}}^l]^{1/3}=1.040$.}
\label{edns_afAfm2}
\end{figure}
As the ${\rm GdFeO}_3$-type distortion increases, the ($yz,xy,xy,zx$)-type
orbital state becomes unstable.
Instead, the solutions with the orbital state
in which $xy$ orbital is mixed into the occupied $yz$ and $zx$ 
orbitals [($yz,zx,yz,zx$)-type orbital state] become stable
(see Fig.~\ref{edns_afAfm2}).
By moderately increasing the distortion, AFM(A) state with 
($yz,zx,yz,zx$)-type orbital ordering is stabilized relative to
FM1 solution.
With further decreasing of the bond angle, the FM state with
($yz,zx,yz,zx$)-type orbital ordering (FM2 solution) is stabilized. 
The AFM(G) solution with ($yz,zx,yz,zx$)-type orbital ordering
[AFM(G)2] has much higher energy relative to the other solutions. 
The AFM(A) to FM2 phase transition occurs
at $\angle$Ti-O-Ti$\sim142^{\circ}$.
These AFM(A) and FM2 states are expected to be realized in the 
systems which are located near the AFM-FM phase boundary.  
In addition, we note that in the large JT distortion of 1.040, the
spin-orbit ground state does not have any stable solutions.

\begin{figure}[tdp]
\hfil
\includegraphics[scale=0.5]{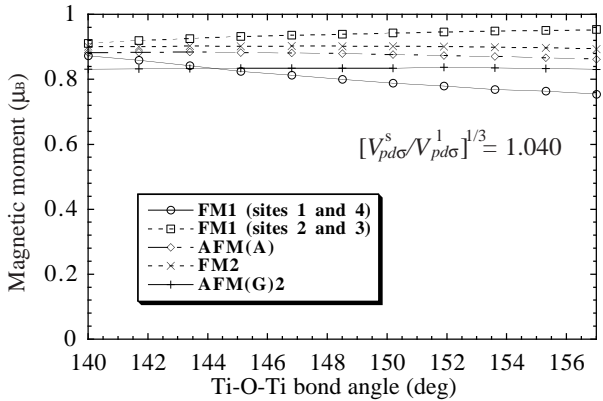}
\hfil
\caption{Magnetic moment of various spin and orbital states
as functions of the Ti-O-Ti bond angle in the case 
of $[V_{pd{\sigma}}^s$/$V_{pd{\sigma}}^l]^{1/3}=1.040$.
In the FM1 state, the magnitude of the magnetic moment 
is different between sites 1, 4 and sites 2, 3.}
\label{mag1.040}
\end{figure}
In Fig.~\ref{mag1.040}, we have plotted the magnetic moment
of the various spin and orbital solutions as functions of the 
Ti-O-Ti bond angle.
With this large JT distortion, 
the orbital angular momentum is mostly quenched and the magnetic 
moment basically consists of the spin-only moment.
This indicates that the effect of the spin-orbit interaction
can be neglected and the system near the phase boundary
is well described by the JT ground state.

Under this large $d$-type JT distortion,
the occupation of the higher $t_{2g}$ orbitals 
at sites 1, 2, 3 and 4 are
close to zero so that the occupied orbitals at each site 
can be expressed by the linear combination of the twofold 
degenerate lowered orbitals, approximately.
In addition, since the order of the indirect $d$-$d$-transfers
mediated by the O $2p$ orbitals is 
$\frac{V_{pd\pi}^2}{\Delta} \sim $ 0.2 eV 
and sufficiently small compared with $U$, the $\vct{$k$}$-dependence
of the coefficients for the linear combinations can be neglected.
So that, we can express the occupied orbitals at $i$-th
site by the coefficients ${C}_{\alpha,i \gamma \sigma}$ as,
\begin{equation}
  {\sum_{\gamma}}^{\prime} {C}_{\alpha,i \gamma \sigma}|\gamma>. 
\label{oo}
\end{equation} 
Here, ${\sum_{\gamma}}^{\prime}$ denotes the summation over the twofold
lowered orbitals in the $d$-type JT distortion at $i$-th site,
namely, $xy$ and $yz$ orbitals at sites 1 and 3, and $xy$ and $zx$ 
orbitals at sites 2 and 4. 
Since the spin-orbit interaction is not effective 
under the large JT distortion, the imaginary parts of the coefficients
are negligible. 
At this stage, we define the absolute values of
the coefficients in the normalized form as,
\begin{equation}
  |{C}_{\alpha,i \gamma \sigma}|=
\sqrt{
\frac{\langle d_{\alpha,i \gamma \sigma}^{\dagger}
              d_{\alpha,i \gamma \sigma} \rangle  }
     {{\sum_{\gamma'}}^{\prime}
      \langle d_{\alpha,i \gamma' \sigma}^{\dagger}
              d_{\alpha,i \gamma' \sigma} \rangle}}.     
\label{coeff}
\end{equation} 

\begin{figure}[h]
  \hfil
\includegraphics[scale=0.5]{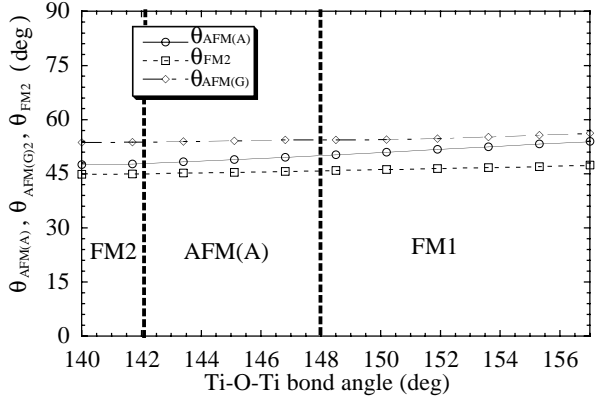}
  \hfil
  \caption{The orbital structures in AFM(A), AFM(G)2 and FM2
solutions as functions of the Ti-O-Ti bond angle.
The difference between those of AFM(A) and FM2 is 
considerably small, particularly for the 
large ${\rm GdFeO}_3$-type distortion or near the AFM(A)-FM2
phase boundary.}
  \label{thetax}
\end{figure}

The orbital states realized in the AFM(A), AFM(G)2 and FM2
solutions can be specified by using angles $\theta_{\rm AFM(A)}$,
$\theta_{\rm AFM(G)2}$ and $\theta_{\rm FM}$ as,
\begin{eqnarray}
&{\rm site}&\quad 1; \quad\cos{\theta_x}|xy>+\sin{\theta_x}|yz>,
\nonumber \\
&{\rm site}&\quad 2; \quad\cos{\theta_x}|xy>+\sin{\theta_x}|zx>,
\nonumber \\
&{\rm site}&\quad 3; \quad -\cos{\theta_x}|xy>+\sin{\theta_x}|yz>,
\nonumber \\
&{\rm site}&\quad 4; \quad -\cos{\theta_x}|xy>+\sin{\theta_x}|zx>,
\end{eqnarray}
where $x =$ AFM(A), AFM(G)2 and FM2.
In Fig.~\ref{thetax}, the angles for the AFM(A), AFM(G)2 and FM2 solutions 
are plotted as functions of the Ti-O-Ti bond angle.
In AFM(A) state, the sites 1, 2, 3 and 4 are occupied by
$c_1yz+c_2xy$, $c_1zx+c_2xy$, $c_1yz-c_2xy$ and $c_1zx-c_2xy$
($c_1^2 + c_2^2 = 1$), respectively.
In particular, the difference between $c_1$ and $c_2$ tends to be small
with increasing the ${\rm GdFeO}_3$-type distortion, and both $c_1$ and $c_2$ 
take approximately $1/\sqrt{2}$ for the large distortion.
Moreover, the similar orbital state is also realized in FM2 state,
and both $c_1$ and $c_2$ also take approximately $1/\sqrt{2}$.
This orbital state is in agreement with previous theoretical 
predictions,~\cite{Mizokawa96a,Mochizuki01a,Sawada97,Sawada98} 
and is observed experimentally in 
${\rm YTiO}_3$.~\cite{Itoh97,Itoh99,Ichikawa00,Akimitsu01}
The difference between $\theta_{\rm AFM(A)}$ and $\theta_{\rm FM2}$
is very small, especially in the largely distorted region or near
the AFM(A)-FM2 phase boundary.
This indicates that the orbital ordering hardly
changes through the magnetic phase transition.
Then, the AFM(A)-to-FM2 phase transition is identified as
a nearly continuous one with a tiny jump in the
spin-exchange interaction along the $c$-axis
from positive to negative and it takes approximately
zero at the phase boundary.
On the contrary, the FM spin-exchange interaction 
is constantly realized in the $ab$-plane.
The resultant strong two-dimensionality in the spin coupling
can cause the strong suppression of $T_{\rm N}$ and $T_{\rm C}$
near the phase boundary.
These are all in agreement with the previous
strong coupling studies~\cite{Mochizuki00,Mochizuki01a} 
and indicate that these results 
are valid even at a realistic and intermediate coupling strength. 

We can also specify the orbital state realized in the 
FM1 solution by using two angles $\theta_1$ and $\theta_2$
as follows,
\begin{eqnarray}
&{\rm site}&\quad 1; \quad\cos{\theta_1}|yz>+\sin{\theta_1}|xy>,
\nonumber \\
&{\rm site}&\quad 2; \quad\cos{\theta_2}|zx>+\sin{\theta_2}|xy>,
\nonumber \\
&{\rm site}&\quad 3; \quad -\cos{\theta_2}|yz>+\sin{\theta_2}|xy>,
\nonumber \\
&{\rm site}&\quad 4; \quad -\cos{\theta_1}|zx>+\sin{\theta_1}|xy>.
\label{eqn:eqtheta1}
\end{eqnarray}

\begin{figure}[tdp]  
\hfil
\includegraphics[scale=0.5]{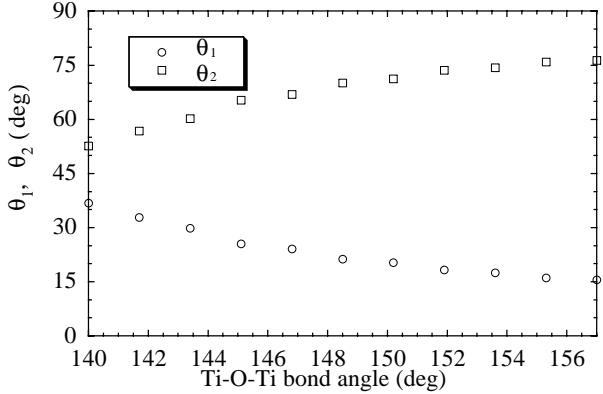}
\hfil
\caption{The orbital structure in the FM1 solution
as a function of the Ti-O-Ti bond angle.
In the small ${\rm GdFeO}_3$-type distortion, an almost complete
($yz,xy,xy,zx$)-type orbital ordering is realized.}
\label{theta_fm1}
\end{figure}
In Fig.~\ref{theta_fm1}, the angles $\theta_1$ and $\theta_2$ are plotted 
as functions of the Ti-O-Ti bond angle.
In the small ${\rm GdFeO_3}$-type distortion 
($\angle$Ti-O-Ti$\sim 157^{\circ}$),
almost complete ($yz, xy, xy, zx$)-type occupation is realized.
In this orbital ordering, the neighboring occupied-orbitals are 
approximately orthogonal and electron transfers from the occupied
orbitals are restricted to neighboring unoccupied orbitals.
However, with increasing the ${\rm GdFeO_3}$-type distortion, 
the occupations of the $xy$, $zx$, $yz$ and $xy$
orbitals gradually increase at sites 1, 2, 3 and 4, respectively
(see also Fig.~\ref{edns_fm1}).
Therefore, both $\theta_1$ and $\theta_2$ 
tend to become close to $45^{\circ}$, and 
the orbital state in the FM1 solution become similar
to that in the FM2 solution as the ${\rm GdFeO_3}$-type distortion
increases.
As a result, FM1 solution in the large ${\rm GdFeO_3}$-type distortion
is similar to that of FM2 so that 
the energy difference between FM1 and FM2
solutions is small in the largely distorted region.
This indicates that the ($yz,zx,yz,zx$)-type orbital ordering realized
in the AFM(A) and FM2 states is strongly stabilized for the
large ${\rm GdFeO_3}$-type distortion.
In addition, we note that with the large JT distortion of 1.040, the
spin-orbit ground state does not have any stable solutions.

\begin{figure}[tdp]
\hfil
\includegraphics[scale=0.5]{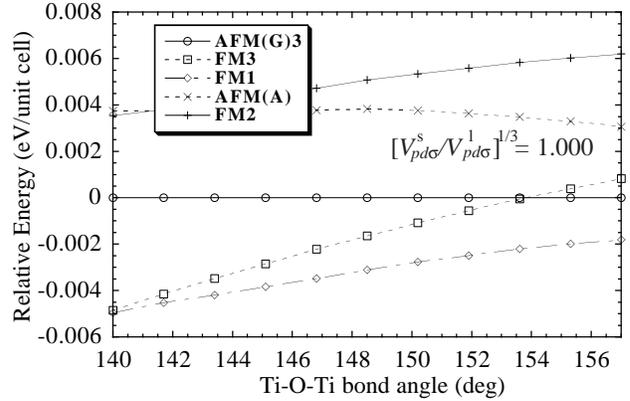}
\hfil
\caption{Energies of various spin and orbital configurations
relative to that of AFM(G)3 state
as functions of the Ti-O-Ti bond angle in the case 
of $[V_{pd{\sigma}}^s$/$V_{pd{\sigma}}^l]^{1/3}=1.000$.}
\label{re1.000}
\end{figure}

We next fix the magnitude of the JT distortion: 
$[V_{pd{\sigma}}^s$/$V_{pd{\sigma}}^l]^{1/3}$
at 1.000 (i.e. no JT distortion) in order to focus on the 
situation realized in ${\rm LaTiO}_3$.
In Fig.~\ref{re1.000}, relative energies of various 
spin and orbital configurations are plotted as functions 
of the Ti-O-Ti bond angle.
Without the JT distortion, AFM(G)1 and AFM(G)2 
states have no stable solutions. 

So far, in the small ${\rm GdFeO}_3$-type distortion, 
the AFM state with spin-orbit ground state [AFM(G)3], 
out of which two states with antiparallel spin and orbital moment,
$\frac{1}{\sqrt{2}}(yz+izx)\uparrow$ and
$\frac{1}{\sqrt{2}}(yz-izx)\downarrow$ 
are alternating between nearest neighbors, 
is considered to be stabilized both
by the spin-orbit interaction and by the superexchange interactions.
However, though this AFM(G)3 state is lower in energy
relative to the FM3 state in which $\frac{1}{\sqrt{2}}(yz+izx)\uparrow$ 
($\frac{1}{\sqrt{2}}(yz-izx)\downarrow$) is occupied
at each site, a FM state with AF-orbital ordering (FM1) is always
lower in energy as compared with AFM(G)3 and FM3 solutions.
This indicates that spin-orbital superexchange interactions caused by 
electron transfers dominate over 
the couplings of the spin and orbitals due to 
the spin-orbit interaction, and the spin-orbit interaction does
not play a role in the emergence of AFM(G) state in ${\rm LaTiO}_3$.  

In the FM1 state, the sites 1, 2, 3, and 4 are approximately
occupied by $\frac{1}{\sqrt{2}}(yz+izx)\uparrow$, $xy\uparrow$, 
$xy\uparrow$ and
$\frac{1}{\sqrt{2}}(yz+izx)\uparrow$, respectively.
At sites 1 and 4, the spin-orbit ground state 
with antiparallel spin and orbital moment is realized.
Since the neighboring $\frac{1}{\sqrt{2}}(yz+izx)\uparrow$ and
$xy$ are approximately orthogonal, AF-orbital ordering 
accompanied by the spin-orbit ground state is realized
in FM1 solution.
Consequently, this FM1 state is strongly stabilized
both by the spin-orbit interaction and by the spin-orbital 
superexchange interactions.

\begin{figure}[tdp]
\hfil
\includegraphics[scale=0.5]{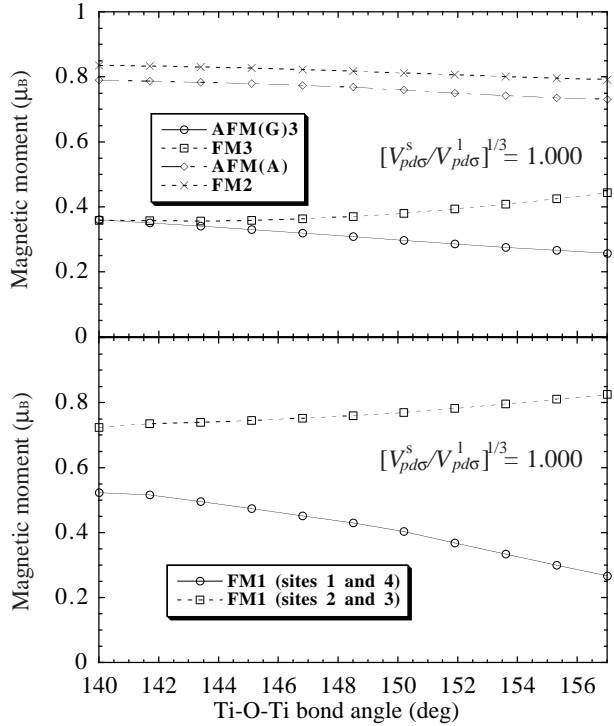}
\hfil
\caption{Magnetic moment of various spin and orbital states
as functions of the Ti-O-Ti bond angle in the case 
of $[V_{pd{\sigma}}^s$/$V_{pd{\sigma}}^l]^{1/3}=1.000$.
In the FM1 state, the magnitude of the magnetic moment 
is different between sites 1, 4 and sites 2, 3.}
\label{mag1.000}
\end{figure}
In Fig.~\ref{mag1.000}, the magnetic moment
of various spin and orbital solutions are
plotted as functions of the bond angle.
In AFM(G)3 and FM3 with spin-orbit ground state, 
the magnetic moment is strongly reduced from
the spin-only moment due to the antiparallel contribution
of the unquenched orbital moment
while those of AFM(A) and FM2 with JT ground state
take approximately unreduced values.
In FM1, reduced ordered moment is realized at sites 1 and 4
with the spin-orbit ground state while the moments are not so reduced   
at sites 2 and 3 in which $xy$ orbital is dominantly
occupied.
 
The strong stabilization of the FM1 state in the small
${\rm GdFeO}_3$-type distortion with no JT distortion
indicates that the spin and orbital states in ${\rm LaTiO}_3$
can not be described by the spin-orbit ground state. 
In addition, there exists a discrepancy between the calculated
energy difference and that expected from experimentally 
obtained $T_{\rm N}$ of $\sim$130 K.
We expect the energy difference between
FM and AFM(G) solutions per unit cell from $T_{\rm N}$
in the following way.
First, we can naively estimate the spin-exchange constant
$J$ in ${\rm LaTiO}_3$ based on a comparison
of $T_{\rm N}$ with the numerical study on the 
spin-1/2 Heisenberg model on a cubic lattice as 
$J=k_{\rm B}T_{\rm N}/0.946 \sim$ 12 meV.~\cite{Sandvik99}
Then, a bond-energy difference between FM and AFM spin configurations
per Ti-Ti bond is $J$/2, and there are 12 Ti-Ti bonds
in the unit cell so that we can expect that AFM(G) solution is
lower than FM solution in energy by $6J\sim72{\rm meV}$
within the Hartree-Fock approximation.
However, this value is considerably large as compared with the
characteristic order of the calculated energy difference
even if the spin-orbit ground state is realized in 
${\rm LaTiO}_3$. (For instance, the energy difference between 
AFM(G)3 and FM3 is $\sim$ 1 meV per unit cell.)
This discrepancy can not be explained within the error
bars of the parameters estimated from the analyses
of photoemission spectra so that we can conclude the spin-orbit 
interaction is irrelevant to the AFM(G) state in ${\rm LaTiO}_3$.

Here, a question arises: why is the ordered  moment reduced 
from $1$ $\mu_{\rm B}$ so strongly if the spin-orbit interaction
can not be its origin?
Recent optical measurement shows
that ${\rm LaTiO}_3$ has a considerably small optical gap of $\sim 0.1$ eV
in the vicinity of the metal-insulator (M-I) phase boundary
with strong itinerant character.~\cite{Okimoto95}
Therefore, in this system,
we expect that some amount of charge and spin fluctuations remain.
The reduction of the magnetic moment may easily be attributed
to this itinerant fluctuation.~\cite{White89}
For instance, in 2D case, 
the ordered moment $\sim 0.6$ ${\mu}_{\rm B}$ for the Heisenberg 
model diminishes to $\le0.2$ ${\mu}_{\rm B}$ 
for $U=4$ Hubbard model due to the itinerant fluctuation 
accompanied by the double occupancy.~\cite{White89}
This strong reduction of the ordered moment with
charge fluctuations is also obtained for Hubbard model 
with next-nearest neighbor transfers in recent 
numerical study.~\cite{Kashima01} 
Within the Hartree-Fock approximation, the ordered moment
is equivalent to the local moment so that the reduction of
the moment can not be reproduced.
However, we can expect this reduction irrespective of the dimensionality
in an insulator with small insulating gap near the 
M-I phase boundary. 
Consequently, though the spin moment within the spin-wave theory 
takes $\sim0.844$ ${\mu}_{\rm B}$ and the reduction due to
the quantum effects is small in 3D spin-wave approximation,
the ordered moment would easily 
diminishes to $\sim0.45$ ${\mu}_{\rm B}$
in ${\rm LaTiO}_3$ with the strong itinerant character and 
large expectation value of the double occupancy
when charge fluctuations are properly taken into account.

\begin{figure}[tdp]
\hfil
\includegraphics[scale=0.5]{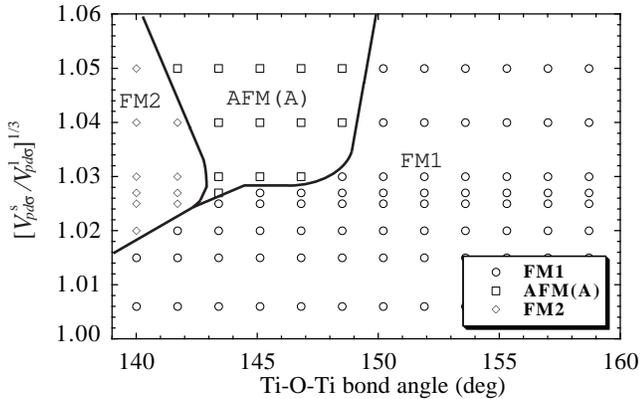}
\hfil
\caption{Magnetic and orbital phase diagram
in the plane of the ${\rm GdFeO}_3$-type and $d$-type JT distortions.}
\label{mpd_dot}
\end{figure}
In Fig.~\ref{mpd_dot}, we show the magnetic and orbital phase diagram
in the plane of the ${\rm GdFeO}_3$-type and $d$-type JT distortions.
In the region of $[V_{pd{\sigma}}^s$/$V_{pd{\sigma}}^l]^{1/3}>1.027$,
AFM(A)-FM phase transition occurs as increasing the
${\rm GdFeO}_3$-type distortion.
In the small ${\rm GdFeO}_3$-type distortion, FM state with
AF-orbital ordering is stabilized in the whole
range of $[V_{pd{\sigma}}^s$/$V_{pd{\sigma}}^l]^{1/3}$.
In particular, in the small JT distortion, only FM1 state
is stabilized.
In FM1 state with no JT distortion, sites 1, 2, 3 and 4 are occupied by
$\frac{1}{\sqrt{2}}(yz+izx)\uparrow$, $xy\uparrow$, $xy\uparrow$
and $\frac{1}{\sqrt{2}}(yz+izx)\uparrow$, respectively, and 
though the spin-orbit ground state is realized at sites 1 and 4, 
$xy\uparrow$-occupancy is favored at sites 2 and 3 by the 
spin-orbital superexchange interactions.
In addition, AFM(G) phase does not exist even for the
small JT distortion.

\section{Summary and Conclusions}

In this paper, we have studied the magnetic and orbital
states and their phase transitions of the perovskite-type 
Ti oxides by using the multiband $d$-$p$ Hamiltonian.
In this Hamiltonian, effects of both electron transfers and 
spin-orbit interaction are considered.
By applying the unrestricted Hartree-Fock approximation to this
Hamiltonian, we have investigated the orbital-spin
states as functions of the magnitudes of 
${\rm GdFeO}_3$-type and $d$-type JT distortions
from a viewpoint of competitions of the spin-orbit interaction,
JT level-splitting and spin-orbital superexchange interactions.
These competitions are characteristic in $t_{2g}$ systems
such as titanates in contrast with $e_g$ systems 
such as manganites since the spin-orbit interaction 
strongly affects the $t_{2g}$ orbitals
relative to the $e_g$ orbitals and JT coupling is rather
weak in $t_{2g}$ systems while 
the coupling almost always dominates over the spin-orbit
interaction in $e_g$ systems.
Our model and approach which treat the electron transfers
and the spin-orbit interaction on an equal footing
and in a non-perturbative manner are appropriate
for the study of the competitions.
We expect that the physics of AFM or AF-orbital 
ordering with superexchange mechanism
in the strong-coupling region is connected   
adiabatically with the SDW-type symmetry breaking
in the weak-coupling Hartree-Fock solutions. 
So that, we have referred the stabilization of the SDW
(or orbital density wave) type  
solution with AFM (or AF-orbital) symmetry breaking to the 
superexchange mechanism.

In the perovskite-type Ti oxides, the transfers of electrons on Ti 
$3d$-orbitals are governed by supertransfer processes 
mediated by the O $2p$ states.
We can calculate the nearest-neighbor and 
next-nearest-neighbor $d$-$d$ transfers
($t$ and $t'$, respectively) by using perturbational 
expansions with respect to $d$-$p$ and $p$-$p$ transfers 
which are determined by using Slater-Koster parameters.
A tight-binding (TB) Hamiltonian with thus obtained $t$ and $t'$
well reproduces the band structure obtained 
in LDA calculations.~\cite{Hamada02}
The characteristic perturbational processes for $t$ and $t'$ are
mediated by one O ion and by two O ions, respectively.
The order of $t$ and $t'$ are $t_{pd}^2/{\Delta}$ and
$t_{pd}^2t_{pp}/{\Delta}^2$ with $t_{pd}$ and $t_{pp}$
being characteristic $d$-$p$ and $p$-$p$ transfers, respectively.
In these compounds, the order of $t_{pp}/{\Delta}$ is about 
$\sim 0.05$ at most so that $t'$ is much smaller than $t$.
Actually, the band structure calculated by using TB
model with both $t$ and $t'$ is almost the same as 
that obtained by using the model with only $t$, particularly
in the $t_{2g}$-band dispersions.
When $t'$ is negligible as compared with $t$, 
we can expect that considerable degree of 
nesting remains. Consequently, in these system, the Hartree-Fock 
calculation can give reliable results for the AFM and AF-orbital type 
symmetry breaking.

Similar weak coupling approach has already been applied to both
end compounds ${\rm LaTiO}_3$ and ${\rm YTiO}_3$
by Mizokawa and Fujimori.~\cite{Mizokawa95,Mizokawa96a}
On the other hand, in this paper, the systems located near the 
AFM-FM phase boundary are studied by this method for the first time.
Moreover, by studying a FM state with lower energy which they
overlooked, we conclude that the spin-orbit interaction can not be an 
origin for the AFM(G) state in ${\rm LaTiO}_3$ in contrast with 
their conclusion. The conclusions of this paper are as follows.

In the region of large JT distortion, the spin-orbit interaction
is dominated by the JT level-splitting and the system is well
described by the JT ground state.
In this region, the AFM(A)-to-FM phase transition 
occurs with increasing the ${\rm GdFeO}_3$-type distortion.
Through this phase transition, the orbital state 
changes negligibly in agreement with the previous 
strong coupling studies.~\cite{Mochizuki00,Mochizuki01a}
The negligible change in the orbital state through this 
AFM(A)-FM phase transition 
causes a nearly continuous change in the spin-coupling
along the $c$-axis, and we can attribute 
the strong suppressions of $T_{\rm N}$ and $T_{\rm C}$
to the resultant two-dimensionality in the spin coupling 
near the phase boundary. 
The orbital states obtained in the FM2 and AFM(A) solutions
are in agreement with those obtained by the previous strong
coupling approaches,~\cite{Mochizuki00,Mochizuki01a} 
which indicates the validity of the results
even at a realistic and intermediate coupling strength. 
Actually, this orbital state has already been observed in 
${\rm YTiO}_3$.~\cite{Itoh97,Itoh99,Ichikawa00,Akimitsu01}
We expect that a similar orbital ordering may be observed in the 
compounds near the AFM-FM boundary 
such as ${\rm SmTiO}_3$, ${\rm GdTiO}_3$ and 
${\rm La}_{1-y}{\rm Y}_{y}{\rm TiO}_3$ ($y\sim0.3$).
Recent resonant x-ray scattering study shows that
the orbital states in ${\rm SmTiO}_3$ and ${\rm GdTiO}_3$
have twofold symmetry similarly to ${\rm YTiO}_3$,
and this seems to be in agreement with our result.~\cite{Kubota}
Here, we note that neutron scattering experiment reveals
that the magnetic structure of the Ti sites
in ${\rm SmTiO}_3$ is not AFM(A) but AFM(G).~\cite{Amow98}
In addition, though ${\rm SmTiO}_3$ is located near the phase 
boundary, $T_{\rm N}$ of $\sim 50$ K is somewhat high relative to the 
previous theoretical prediction.~\cite{Mochizuki00,Mochizuki01a} 
In ${\rm SmTiO}_3$, there exist magnetic moments on the Sm sites,
and Sm-Ti spin-coupling may be important for its magnetic properties
while our model does not take the orbital and spin degrees 
of freedom on the $R$ sites into account.
However, our model can well describe the orbital-spin states
and their phase transitions of ${\rm LaTiO}_3$, ${\rm YTiO}_3$
and ${\rm La}_{1-y}{\rm Y}_{y}{\rm TiO}_3$
systems with no magnetic moments on La and Y sites.
Moreover, since the orbital state near the AFM-FM phase boundary 
is strongly stabilized irrespective of the spin structure
as shown in both our weak-coupling and previous strong-coupling 
studies, the similar $(yz,zx,yz,zx)$-type orbital state
is also expected to be realized in ${\rm SmTiO}_3$
though the magnetic structure is AFM(G) due to the 
Sm-Ti spin-coupling.

Without a JT distortion, owing to both 
spin-orbit and spin-orbital superexchange interactions 
a FM state with the spin-orbit ground state 
accompanied by an AF-orbital ordering 
[($\frac{1}{\sqrt{2}}(yz+izx)\uparrow$, $xy\uparrow$, $xy\uparrow$,
$\frac{1}{\sqrt{2}}(yz+izx)\uparrow$)-orbital ordering]
is stabilized relative to the other solutions.
This FM solution can not be reproduced by the previous strong coupling 
approach in which the spin-orbit interaction is neglected in the large 
JT distortion, and is studied for the first time by our 
weak coupling approach. 
In addition, AFM(G) state is higher in energy and has no
stable solutions.
While in this system, the spin-orbit interaction 
has been considered to be relevant
in the small or no JT distortion so far, the spin-orbital superexchange
interactions due to the electron transfers 
turn out to dominate over the spin-orbit interaction.  
Moreover, if we would take the dominance of the
spin-orbit interaction, there would be 
a discrepancy between the calculated energy-difference
and that estimated from $T_{\rm N}$.
Thus, we conclude that the spin-orbit interaction is irrelevant to the origin
of AFM(G) state in ${\rm LaTiO}_3$, and the experimentally observed
reduction of the moment can be attributed to the strong itinerant
fluctuations in ${\rm LaTiO}_3$ instead of the spin-orbit interaction.
Indeed, a recent neutron-scattering experiment
reveals the spin-wave spectrum of ${\rm LaTiO}_3$
well described by a spin-1/2 isotropic Heisenberg model
on the cubic lattice and the absence of unquenched 
orbital angular momentum.~\cite{Keimer00}
This indicates that the spin-orbit interaction is not 
effective in this system.
Our results support these experimental results and suggest another 
mechanism for the emergence of the AFM(G) state.
We expect that effects which are not treated in our model 
are responsible for its origin.
Recently, possible $D_{3d}$ distortion of the ${\rm TiO}_6$ octahedron
is examined as a candidate for the origin and nature of AFM(G) state
in ${\rm LaTiO}_3$.~\cite{Mochizuki01b} 
In this scenario, the spin-orbit interaction is dominated 
by the $t_{2g}$-level splitting due to the $D_{3d}$ crystal field.

In addition, we have also studied a magnetic phase diagram
in the plane of the ${\rm GdFeO}_3$-type and $d$-type JT distortions
in order to examine how extent the physics of AFM-FM phase transition
in strong coupling limit survives when the JT level-splitting
competes with the spin-orbit interaction.
According to the obtained phase diagram, the 
description of the phase transition obtained by the previous strong 
coupling approach is well established in the wide range
of JT distortion even when the spin-orbit interaction is
taken into consideration.

\section*{Acknowledgement}

The author would like to thank M. Imada, T. Mizokawa, 
H. Asakawa, Y. Motome and N. Hamada for valuable discussions
and useful comments. 
This work is supported by ``Research for the Future Program''
(JSPS-RFTF97P01103) from the Japan Society for the Promotion of Science.



\end{document}